\documentclass{article}
\usepackage[T1]{fontenc}
\usepackage[utf8]{inputenc}
\usepackage{ismir} 
\usepackage{amsmath,cite,url}

\usepackage{graphicx}
\usepackage{color}
\usepackage[margin=1in]{geometry}
\usepackage{array}
\usepackage{xcolor}
\usepackage{booktabs}
\usepackage{amssymb}
\usepackage{amsmath}
\usepackage{multirow}
\usepackage{algorithm}
\usepackage{algpseudocode}

\title{Count The Notes: Histogram-Based Supervision for Automatic Music Transcription}

\multauthor
  {Jonathan Yaffe$^1$ \hspace{1cm} Ben Maman$^2$ \hspace{1cm} Meinard Müller $^2$ \hspace{1cm} Amit H. Bermano$^1$}
  {
    $^1$ Tel Aviv University, Israel\\
  $^2$ International Audio Laboratories Erlangen\\
  {\tt\small jonathany@mail.tau.ac.il, ben.maman@audiolabs-erlangen.de}
  }

\def\authorname{J. Yaffe, B. Maman, M. Müller, and A.H. Bermano}

\usepackage[bookmarks=false,pdfauthor={\authorname},pdfsubject={\pdfsubject},hidelinks]{hyperref}

\begin{document}

\maketitle
\begin{abstract}
Automatic Music Transcription (AMT) converts audio recordings into symbolic musical representations. Training deep neural networks (DNNs) for AMT typically requires strongly aligned training pairs with precise frame-level annotations. Since creating such datasets is costly and impractical for many musical contexts, weakly aligned approaches using segment-level annotations have gained traction. However, existing methods often rely on Dynamic Time Warping (DTW) or soft alignment loss functions, both of which still require local semantic correspondences, making them error-prone and computationally expensive. In this article, we introduce CountEM, a novel AMT framework that eliminates the need for explicit local alignment by leveraging note event histograms as supervision, enabling lighter computations and greater flexibility. Using an Expectation-Maximization (EM) approach, CountEM iteratively refines predictions based solely on note occurrence counts, significantly reducing annotation efforts while maintaining high transcription accuracy. Experiments on piano, guitar, and multi-instrument datasets demonstrate that CountEM matches or surpasses existing weakly supervised methods, improving AMT's robustness, scalability, and efficiency. Our project page is available at \href{https://yoni-yaffe.github.io/count-the-notes}{\url{https://yoni-yaffe.github.io/count-the-notes}}
\end{abstract}
\section{Introduction}\label{sec:introduction}
\begin{figure*}[t!]
    \centering
    \includegraphics[width=\textwidth]{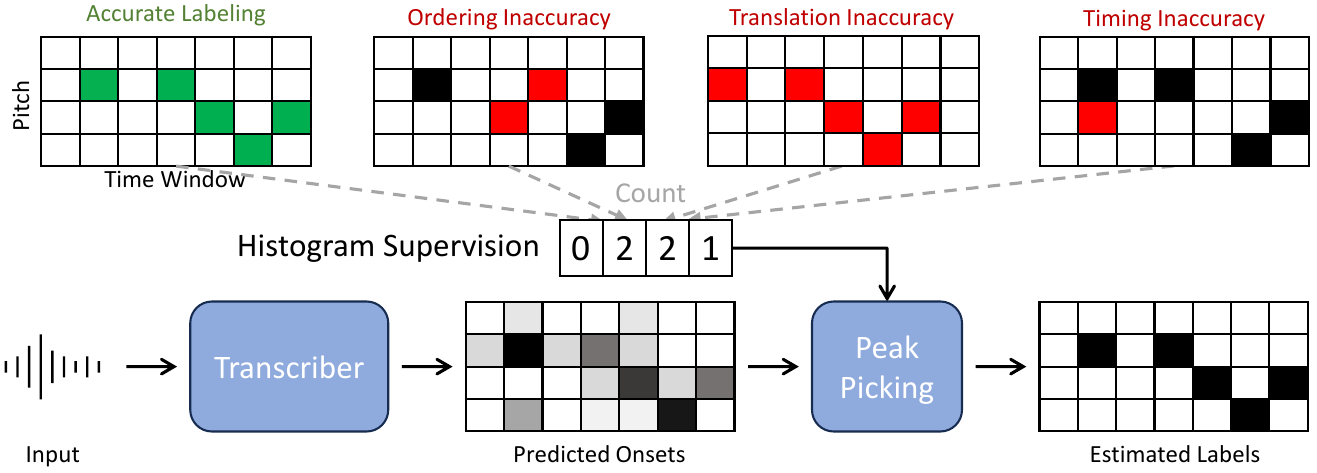}
    \caption{
    Estimating aligned labels from histograms by peak-picking. 
    For each note in the histogram, the $K$ most likely timings are selected according to the current predicted posteriorgram.
    Since misaligned labels reduce to the same histogram (top), possible timing inaccuracies common in weakly-aligned labels can be overcome.
    }

    \label{fig:labeling}
\end{figure*}
Automatic Music Transcription (AMT) converts audio recordings into symbolic, score-like representations. As a core task in Music Information Retrieval (MIR), AMT has applications in music education, analysis, production, and neural generation. However, it remains challenging, particularly for polyphonic and multi-instrument recordings, due to overlapping harmonics, complex timbres, and varying acoustic environments. Most AMT systems rely on strongly aligned training data, where each audio frame has an exact corresponding label~\cite{HawthorneESRSRE18_OnsetsFrames_ISMIR, HawthorneSRSHDE19_MAESTRO_ICLR, KongEtAl21_HighResTranscription_TASLP, GardnerSMHE22_MultiTaskTranscription_ICLR}. While effective, creating such datasets is costly and labor-intensive, restricting AMT models to specific instruments, styles, and acoustic conditions. As an alternative, semi-supervised learning methods use weakly aligned segment-level annotations rather than frame-level labels,
showing that imperfect supervision---
such as unaligned
transcriptions from different performances of the same piece,
can still provide useful training targets~\cite{MamanBermano22_UnalignedAMT_ICML,RileyED24_GuitarTranscription_ICASSP, RileyGED2024_GAPS_ISMIR, ZalkowMueller21_ChromaCTC_TASLP}.

One such method, NoteEM~\cite{MamanBermano22_UnalignedAMT_ICML}, applies an Expectation-Maximization (EM) framework to iteratively refine weak labels. Beginning with a transcriber trained on synthetic data, it alternates between aligning weak labels using the network's predicted features, and training the network with these labels. This strategy has achieved high transcription accuracy across diverse musical styles and instruments~\cite{MamanBermano22_UnalignedAMT_ICML,RileyED24_GuitarTranscription_ICASSP, RileyGED2024_GAPS_ISMIR}.
However, alignment methods like Dynamic Time Warping (DTW)~\cite{Mueller07_InformationRetrieval_SPRINGER}
introduce synchronization errors, computational overhead, and label inconsistencies,
even with improved neural features. This is especially true for note onset detection, where high temporal precision is crucial~\cite{HawthorneESRSRE18_OnsetsFrames_ISMIR, HawthorneSRSHDE19_MAESTRO_ICLR, MamanBermano22_UnalignedAMT_ICML, RileyED24_GuitarTranscription_ICASSP}. Most critically, such approaches assume weak labels preserve event order, even if misaligned---an assumption that often fails in real-world scenarios, such as in arpeggios, where chords are performed as sequential notes.

As the main contribution of this article, we introduce \textbf{CountEM}, a novel AMT framework leveraging an even weaker form of supervision: note event counting, integrated with the Expectation-Maximization (EM) algorithm. Unlike supervised or weakly supervised methods that require structural alignment,
CountEM uses
note onset histograms to iteratively refine predictions and temporal estimates. A key insight of CountEM is that strict alignment steps based on approximate temporal ordering, enforced by methods like DTW, can be relaxed or eliminated. Instead of enforcing structure-preserving alignment, CountEM counts note onsets within large time windows, using these counts alone as supervision. This reduces annotation effort while improving efficiency, flexibility, and robustness. Compared to DTW-based methods, histogram-based alignment is computationally simpler,
and minimizes alignment errors caused by structural variations in musical performances.

To demonstrate the effectiveness of CountEM, we adapt the NoteEM framework to our histogram-based supervision approach. The model is initially trained on synthetic data, or other timing-accurate sources, before undergoing an iterative process of labeling and training. During labeling, the model generates onset estimates for each pitch over the prediction temporal window, and the $K$ most probable timings are selected, where $K$ is the supervised event count. See also Figure~\ref{fig:labeling} for an illustration of the process. This method is applicable at various granularities, from entire audio tracks to smaller segments of 30 seconds, with longer windows providing weaker supervision. We evaluate CountEM on real-world datasets, showcasing its ability to generalize across diverse musical contexts, and demonstrate that it matches or surpasses existing weakly supervised methods. Even with large window sizes (up to entire tracks), it maintains high transcription accuracy. Furthermore, we demonstrate CountEM is robust to misalignments and annotation errors, enhancing AMT's scalability and extending its applicability to under-documented musical traditions.

The remainder of this article describes our approach in detail. Section~\ref{sec:method} introduces the methods underlying CountEM, followed by Section~\ref{sec:experiments}, which presents the experiments and evaluation. Section~\ref{sec:conclusion} discusses key findings and implications, with directions for future research. Code and qualitative samples can be found on our project page.\footnote{\href{https://yoni-yaffe.github.io/count-the-notes}{\url{https://yoni-yaffe.github.io/count-the-notes}}}

\section{Method}\label{sec:method}

Note histograms in musical performances can often be accurately derived from sheet music,
particularly for Western classical music, which follows a musical score.
CountEM leverages this information as coarse supervision for music transcription. The central insight of this work is that such counting supervision, which is easy to label and does not require precise timing or note ordering, can be a sufficient training signal. A second insight is that note onsets are prominent features in musical performances and remain consistent between a score and its rendition:
If a note occurs $K$ times in a musical score, then $K$ onsets of that note will be perceived in an actual performance of that score. Indeed, studies on audio--score synchronization demonstrate improved
alignment robustness when incorporating onset features~\cite{EwertMG09_HighResAudioSync_ICASSP, MamanBermano22_UnalignedAMT_ICML, OezerIAM22_ActivationMusicSync_ISMIR, ZeitlerMM24_SYNCHRONIZATION_ISMIR}.

Other performance aspects, such as relative note timing, durations, intensity, and pitch fluctuations, vary by performer and interpretation.
Traditional audio--score synchronization algorithms struggle with these variations, especially in polyphonic music, often leading to alignment errors~\cite{MamanBermano22_UnalignedAMT_ICML, RileyED24_GuitarTranscription_ICASSP}. These errors stem from expressive timing and minor shifts in note order, such as in arpeggios. Effective alignment algorithms, especially for note onsets, must accommodate such variations.
Recent transcription methods use DTW with neural onset features, followed by a refinement step that applies local temporal adjustments for each note independently~\cite{MamanBermano22_UnalignedAMT_ICML, RileyED24_GuitarTranscription_ICASSP}.

In contrast, our method alleviates the need for alignment and DTW by adopting a simpler, more flexible approach. Instead of enforcing strict temporal alignment, we use peak-picking to identify the $K$ most probable onsets in a temporal window based on local maxima in the output signal.
This straightforward, optimization-free process is robust to structural, timing, and ordering inaccuracies. The method follows an EM loop, alternating between label refinement and model improvement (see Section~\ref{sec:e-m}). The E-step refines labels using peak-picking (Section~\ref{sec:labeling}), while the M-step updates network parameters.
\subsection{Expectation--Maximization}\label{sec:e-m}
The EM process, outlined in Algorithm~\ref{algo:em}, consists of the following steps:
\begin{itemize}
    \item \textbf{Initialization:} The model is pre-trained on fully supervised data from an easily accessible domain, such as synthetic data.
    \item \textbf{Expectation (E-step):} 
    The model predicts a note onset posteriorgram (heatmap). The likelihoods in the posteriorgram are refined using top-$K$ local-maxima peak picking for each pitch, based on its target number of occurrences, to estimate strongly-aligned onset labels. As a regularization, we only update the estimated label if the Euclidean distance between the current predicted histogram and the target histogram has improved.
    \item \textbf{Maximization (M-step):} We use the estimated strongly-aligned labels to update the model parameters using standard optimization~\cite{KingmaB15_OptimizerADAM_ICLR}.
\end{itemize}
The E- and M-steps are alternately repeated until convergence. We used 5 iterations for our experiments. The EM iterations progressively improve temporal localization without relying on detailed temporal annotation. Temporal precision is derived from the model itself, which is pre-trained on another domain.


\subsection{Strong Alignment from Histograms}\label{sec:labeling}
\begin{algorithm}[tb]
   \caption{CountEM}
   \label{algo:em}
\begin{algorithmic}
    \State {\bfseries Input:} audio $a_1, \dots a_N$, histog. $h_1,\dots h_N\in\mathbb{N}_0^P$
   \State {\bfseries Output:} model $f_\Theta$, labels $Y_1, \dots Y_N\in\{0, 1\}^{T\times P}$
   \State pre-train $f_\Theta$ (synthetic / other instrument)
   \State $Y_i, d^{\mathrm{hist}}_i = \mathrm{None}, \infty\quad i=1,\dots,N$
   \Repeat
   \For{$i=1$ {\bfseries to} $N$}
   \State $Y_i^{\mathrm{temp}} = \mathrm{PeakPick}(f_\Theta(a_i), h_i)$
   \State $h_i^{\mathrm{pred}} = \Sigma_{t=1}^{T}f_\Theta(a_i)_t\in\mathbb{R}_{+}^P$
   \State $d^{\mathrm{temp}}_i =\|h_i^{\mathrm{pred}} - h_i\|_2$
   \If {$d^{\mathrm{temp}}_i < d^{\mathrm{hist}}_i$}
   \State $Y_i, d^{\mathrm{hist}}_i = Y_i^{\mathrm{temp}}, d^{\mathrm{temp}}_i$
   \EndIf
   \EndFor
   \State $\Theta = \underset{\Theta '}{\mathrm{argmin}}\sum_{i=1}^N \mathrm{BCE}(f_{\Theta '}(a_i), Y_i)$
   \Until{$\sum_{i=1}^N d^{\mathrm{hist}}_i$ converges}
\State \textbf{return} $f_\Theta$, $Y_1, \dots Y_N$
\end{algorithmic}
\end{algorithm}

We use peak-picking to estimate precise time-aligned labels based on the target note histograms and the model's predictions. We assume a target histogram $h=(h_1,\dots,h_P)^\top\in\mathbb{N}_{0}^P$ where $P$ is the number of considered pitches,
and a predicted note onset posteriorgram $Z\in[0,1]^{T\times P}$, where $T$ is the number of time frames. The posteriorgram $Z$ can be interpreted as a predicted note onset
heatmap, which we assume is computed as $Z = f_\Theta(a)$ for a given
input audio representation $a$ and a deep neural network $f_\Theta$.

We assign an estimated label $Y\in[0,1]^{T\times P}$ using a peak-picking operator (``$\mathrm{PeakPick}$'' in Algorithm~\ref{algo:em}):
\begin{equation}
    \Psi:[0,1]^{T\times P}\times\mathbb{N}_0^P\to\{0,1\}^{T\times P}
\end{equation}
which simply picks for each pitch $p \in \{1, \ldots, P\}$ the $K$ most likely temporal local peaks according to the predicted posteriorgram $Z$, where $K=h_p$ is the target number occurrences of the pitch according to the histogram. 
A position is considered to be a local peak if it is higher or equal to all its neighbors in a certain radius of frames, e.g., one frame.

Denoting $Y = \Psi(Z, h)$, for each pitch $p$
the peak picker $\Psi$ selects $K=h_p$ peaks from the $p$-th column of $Z$ to define the $p$-th column of $Y$, where peak positions are binary-encoded (multi-hot).
Note that by definition, it holds that
$\Sigma_{t=1}^{T}Y_t = h\in\mathbb{N}_0^P$,
i.e., the rows of $Y$ sum up to the target histogram $h$.

\subsection{Model Training}
We experiment with two models: The Onsets and Frames architecture~\cite{HawthorneESRSRE18_OnsetsFrames_ISMIR, HawthorneSRSHDE19_MAESTRO_ICLR} pre-trained on synthetic data~\cite{MamanBermano22_UnalignedAMT_ICML}, which we denote $\texttt{Sy}$, and the model of~ Kong et al.\cite{KongEtAl21_HighResTranscription_TASLP} pre-trained on the MAESTRO dataset, which we denote $\texttt{Kg}$.
We
optimize the mean binary cross-entropy (BCE) loss using an Adam optimizer~\cite{KingmaB15_OptimizerADAM_ICLR}. To address the imbalance between positive and negative labels resulting from note onset sparsity,
we assign a
weight \(w \ge 1\) to positive labels during training.
This is done by applying a mask
$M= w\cdot Y + (1-Y)$
to the binary cross-entropy loss matrix,
where $Y$ is the estimated label. The loss function is computed as:
\[
    \mathcal{L}(f_\theta(a), Y) = \sum_{i,j} M_{i, j} \cdot \mathrm{BCE}(f_\theta(a)_{i, j}, Y_{i, j}).
\]
We set the weight \(w\) to 2 ($\texttt{Sy}$) or 1 ($\texttt{Kg}$),
which from our observation provided approximately equal precision and recall.
We apply pitch shift augmentation~\cite{ThickstunHFK18_Transcription_ICASSP, MamanBermano22_UnalignedAMT_ICML, RiouLHP23_PESTO_ISMIR, RileyED24_GuitarTranscription_ICASSP}, generating 11 pitch-shifted copies of the audio data, with shifts in the range of $\pm5$ semitones, and with an additional random fractional term in the range of $\pm 0.1$ semitones to account for small tuning variation. Labels were computed only for the original copy and transposed accordingly for each augmented copy, enforcing pitch shift equivariance. 
All experiments were implemented in PyTorch and executed using two NVIDIA GeForce RTX 3090 GPUs. We used a batch size of 16
and trained models for 37.5K steps, except for Section~\ref{sec:piano}, where we trained for 500K steps. 
\section{Experiments}\label{sec:experiments}
In this section, we present our experiments evaluating our approach across different datasets and instruments, including piano transcription and noisy histograms (Sections~\ref{sec:piano}-\ref{sec:noisy},
MAESTRO dataset~\cite{HawthorneSRSHDE19_MAESTRO_ICLR}),
guitar transcription (Section~\ref{sec:guitar}, cross-dataset), and multi-instrument transcription including strings and winds (Section~\ref{sec:musicnet}, cross-dataset). Evaluation metrics include note-level precision, recall, and F-score with a $50$\,ms onset tolerance.
\subsection{Piano Transcription---MAESTRO Dataset}\label{sec:piano}
\begin{table}[t!]
\centering
\setlength{\tabcolsep}{4pt}
\begin{tabular}{p{0.4cm}c|ccc|ccc}
\hline
\multicolumn{2}{l}{\textbf{Model}} & \multicolumn{3}{c|}{\textbf{Test}} & \multicolumn{3}{c}{\textbf{Train}} 
\\
\cline{3-8}
 \multicolumn{2}{c}{} &  \textbf{P} & \textbf{R} & \textbf{F} & \textbf{P} & \textbf{R} & \textbf{F} \\
\hline
& \multicolumn{7}{c}{Pre-trained Model}\\
\hline
& \texttt{Sy} & 88.3 & 81.6 & 84.6 & 87.8 & 81.2 & 84.1 \\
\hline

& \multicolumn{7}{c}{Histogram Supervision }\\
\hline
\multirow{5}{0.4cm}{\\ Rep. \\ iter. } & \texttt{F/T} & 92.4 & 90.4 & 91.3 & 91.8 & 90.5 & 91.1 \\
& \texttt{180s}   & 93.2 & 91.7 & 92.4 & 92.9 & 91.9 & 92.4 \\
& \texttt{120s}   & 93.1 & 92.2 & 92.6 & 92.8 & 92.4 & 92.6 \\
& \texttt{60s}   & 95.7 & 92.2 & 93.9 & 95.6 & 92.5 & 94.0 \\
& \texttt{30s} & 95.5 & 92.8 & 94.1 & 95.3 & 93.1 & 94.2 \\
\hline
\multirow{2}{0.4cm}{1-iter.}
& \texttt{F/T} & 92.4 & 87.1 & 89.6 & 91.9 & 87.3 & 89.5 \\
& \texttt{60s}   & 93.9 & 88.4 & 91.0 & 93.6 & 88.5 & 90.9 \\
\hline
& \texttt{Sup} & 98.7 & 93.1 & 95.8    & 98.8 & 93.4 & 96.0  \\
\bottomrule
\end{tabular}
\caption{
Note-level transcription results for training with histogram-based supervision on the MAESTRO dataset. We report Precision (P), Recall (R), and F-score (F)
across different histogram window sizes (or Full Track). For reference, results include a  baseline trained on synthetic data only (\texttt{Sy}) and a supervised model trained with ground-truth labels (\texttt{Sup}).
}
\label{table:maestro}
\end{table}

We first evaluate our method in a controlled setting using the MAESTRO dataset~\cite{HawthorneSRSHDE19_MAESTRO_ICLR}, which provides precise reference annotations generated automatically by a Disklavier. Instead of using these labels directly for training, we derive onset histograms by segmenting the audio and labels into smaller windows along the time axis, over which we compute histograms. These histograms serve as supervision for training, while evaluation is performed using the reference labels. To assess the impact of supervision levels, we test window lengths of 30 seconds, one minute, two minutes, three minutes, and entire tracks (up to 40 minutes).

Table~\ref{table:maestro} shows that our approach significantly improves transcription accuracy compared to the initial pre-trained model (\texttt{Sy}), even with full-track histograms (\texttt{F/T}), where F-score increases by over $6\,\%$ (from  $84.6$ to $91.3$). Reducing the counting window further improves the F-score, as it better constrains onset timing, effectively increasing supervision. Performance approaches fully supervised levels for windows of one minute or less, indicating that the counting approach is effective even with temporally highly inaccurate labeling.

We also observe that repeating the labeling process during training (``Rep. iter.'') improves performance compared to training for the \textit{same number of total steps} with a single labeling (``1-iter''), e.g., from $91.0$ to $93.9$ for one-minute windows.

\subsection{Noisy Histograms}\label{sec:noisy}
\begin{table}[t!]
\centering
\setlength{\tabcolsep}{4pt}
\begin{tabular}{lccc|ccc}
\hline 
\textbf{Model}& \multicolumn{3}{c|}{\textbf{Test}} & \multicolumn{3}{c}{\textbf{Train}} \\ 
\cline{2-7}
& \textbf{P} & \textbf{R} & \textbf{F} & \textbf{P} & \textbf{R} & \textbf{F} \\
\hline
\multicolumn{7}{c}{Noisy Histogram Supervision}\\
\hline
\texttt{60s0\%}   & 95.7 & 92.2 & 93.9 & 95.6 & 92.5 & 94.0 \\
\texttt{60s10\%} & 93.1 & 92.0 & 92.5 & 93.0 & 92.2 & 92.6  \\
\texttt{60s20\%} & 92.2 & 90.2 & 91.2 & 91.7 & 90.6 & 91.1 \\
\texttt{F/T0\%}  & 92.4 & 90.4 & 91.3 & 91.8 & 90.5 & 91.1 \\
\texttt{F/T10\%}     & 90.9 & 89.8 & 90.3 & 90.4 & 89.9 & 90.1 \\
\texttt{F/T20\%}     & 89.2 & 88.2 & 88.6 & 88.5 & 88.4 & 88.4 \\
\hline
\end{tabular}
\caption{
CountEM robustness to noisy histograms on the MAESTRO dataset. We apply $\pm$10\% and $\pm$20\% random noise to simulate histogram errors and evaluate different window lengths as in Table~\ref{table:maestro}. 
}
\label{table:model_metrics_noise}
\end{table}
While fully supervised datasets like MAESTRO provide near-perfect histograms, labels for real-world recordings rely on musical scores, introducing potential discrepancies. For example, trills performed differently in audio and unaligned labels can cause minor inconsistencies. To assess the robustness of our approach, we train on the MAESTRO dataset with multiplicative random noise sampled from the uniform distribution $U[1-\alpha, 1+\alpha]$ at two levels ($\alpha \in \{0.1, 0.2\}$), introducing up to 10\% and 20\% noise. We conduct experiments using both one-minute and full-track histograms. Table~\ref{table:model_metrics_noise} shows that while histogram errors slightly affect performance, the impact remains limited---no more than 3\% even with 20\% noise.

\subsection{Guitar Transcription}\label{sec:guitar}
As a next step, we evaluate our method on guitar datasets, namely GuitarSet~\cite{XiBPYB18_GUITARSET_ISMIR} and the Guitar-Aligned Performance Scores (GAPS) dataset~\cite{RileyGED2024_GAPS_ISMIR}. The annotation for GuitarSet was created by applying $f_0$ estimation on monophonic tracks obtained from hexaphonic pickup, followed by semi-automated methods for note onset and offset localization. The annotation for GAPS was done directly on polyphonic tracks by professional annotators, relying on recent neural network-based alignment techniques~\cite{RileyGED2024_GAPS_ISMIR}.

We compare two existing off-the-shelf models: The Onsets and Frames architecture~\cite{HawthorneESRSRE18_OnsetsFrames_ISMIR, HawthorneSRSHDE19_MAESTRO_ICLR} pre-trained on synthetic data~\cite{MamanBermano22_UnalignedAMT_ICML}, and the model of Kong et al.~\cite{KongEtAl21_HighResTranscription_TASLP} pre-trained on the MAESTRO dataset. We denote these models \texttt{Sy} and \texttt{Kg}, respectively. We train each of them using histogram supervision on each of the two datasets---GuitarSet and GAPS, which we denote \texttt{Gs} and \texttt{Gp}, respectively. This yields four different configurations: \texttt{SyGs}, \texttt{SyGp}, \texttt{KgGs}, \texttt{KgGp}. We evaluate each configuration on each of the two datasets, enabling both intra- and inter-dataset (cross dataset) evaluation. We train each of the four configurations with histograms computed over different windows---one-minute windows (\texttt{60s}) and entire tracks (\texttt{F/T}). 

The tracks in GuitarSet are all shorter than 30 seconds, therefore we only use entire-track histograms for it. Since GuitarSet is small (three hours) we train \texttt{SyGs} and \texttt{KgGs} on the entire set, however, only with histogram information. Therefore evaluation of \texttt{SyGs} and \texttt{KgGs} on GuitarSet measures the ability to restore the original time-aligned labels from the histogram information. When training on GAPS, we use the same train--test split as Riley et al.~\cite{RileyGED2024_GAPS_ISMIR}.



Results are shown in Table~\ref{table:guitar_set}. It can be seen that our approach yields significant improvement over both baselines (\texttt{Sy}, \texttt{Kg}) of over $15\%$ in F-score for both GuitarSet and GAPS, even when counting over entire tracks ($\texttt{SyGpF/T}$, $\texttt{KgGpF/T}$). For example, fine-tuning $\texttt{Sy}$ on GAPS with histogram supervision over entire tracks ($\texttt{SyGpF/T}$) improves accuracy on GuitarSet from $66.2\%$ to $84.6\%$.

When reducing the counting window on GAPS to one minute, Accuracy on GuitarSet slightly improves by $1.1\%$ on average.

It can also be seen that
by training on GuitarSet with only its histogram information we can restore its ground-truth strongly-aligned labels with accuracy of $88.9\%$ (\texttt{SyGsF/T}) or $89.7\%$ (\texttt{KgGsF/T}).

We further compare our results to previous work in weakly-supervised transcription. The model of Maman and Bermano~\cite{MamanBermano22_UnalignedAMT_ICML} was fine-tuned from synthetic (the same pre-trained model we use) to self-collected guitar data. We denote this model by \texttt{SySc}. Accuracy of our model surpasses this model, improving on GuitarSet from $82.2\%$ to $85.8\%$ (\texttt{SyGp60s}) or $86.5\%$ (\texttt{KgGp60s}), and on GAPS from $86.6\%$ to $90\%$ (\texttt{SyGp60s}) or $93\%$ (\texttt{KgGp60s}).


The models of Riley et al.~\cite{RileyGED2024_GAPS_ISMIR} were trained on GAPS with its time-aligned labels either from scratch (\cite{RileyGED2024_GAPS_ISMIR}~\texttt{Gp}) or fine-tuned from piano (\cite{RileyGED2024_GAPS_ISMIR}~\texttt{KgGp}). The labels were obtained by alignment of neural onset features, applying an initial DTW step, followed by a local-max refinement step for each note onset independently. Contrary to Riley et al.~\cite{RileyGED2024_GAPS_ISMIR}, we train on GAPS using histogram information only, i.e., weakening or completely omitting the DTW step. Our model's accuracy is slightly higher than~\cite{RileyGED2024_GAPS_ISMIR} trained on GAPS from scratch, and slightly lower than~\cite{RileyGED2024_GAPS_ISMIR} fined tuned on GAPS from MAESTRO,
but on a comparable scale. This shows that the DTW step may be omitted with a small impact.

Most importantly, results show that our approach is robust across different architectures, and enables adaptation to guitar transcription from either synthetic (\texttt{Sy}) or piano (\texttt{Kg}) data pre-training.

\begin{table}[t!]
\centering
\setlength{\tabcolsep}{4pt}
\begin{tabular}{lccc|ccc}
\hline
\textbf{Model}& \multicolumn{3}{c|}{\textbf{GuitarSet}} & \multicolumn{3}{c}{\textbf{GAPS}} \\ 
\cline{2-7}
 & \textbf{P} & \textbf{R} & \textbf{F} & \textbf{P} & \textbf{R} & \textbf{F} \\
\hline
\multicolumn{7}{c}{Pre-trained Models}\\
\hline
\texttt{Sy} & 57.9 & 80.7 & 66.2 & 67.2 & 86.3 & 75.0 \\
\texttt{Kg} & 71.1 & 44.0 & 50.9 & 61.9 & 77.7 & 67.1\\
\hline
\multicolumn{7}{c}{Histogram Supervision}\\
\hline
\texttt{SyGsF/T} & 87.6 & 90.3 & 88.9 & 84.2 & 81.2 & 82.2\\
\texttt{SyGpF/T} & 83.6 & 86.4 & 84.6 & 90.6& 90.6& 90.6\\
\texttt{SyGp60s} & 85.6 & 86.5 & 85.8 & 89.8 & 90.1 & 90.0\\
\texttt{KgGsF/T} & 89.3 & 90.1 & 89.7 & 85.4 & 89.0 & 87.1 \\
\texttt{KgGpF/T}  & 83.6 & 88.1 & 85.5 & 93.3 & 92.5 & 92.9 \\
\texttt{KgGp60s} & 86.9 & 85.4 & 86.5 & 93.1 & 93.0 & 93.0 \\
\hline
\multicolumn{7}{c}{DTW + Refinement}\\
\hline
\cite{RileyGED2024_GAPS_ISMIR} \texttt{Gp} & 92.4 & 81.8 & 86.1 & 94.9 & 92.1 & 93.4\\
\cite{RileyGED2024_GAPS_ISMIR} \texttt{KgGp} & 91.1 & 85.9 & 88.1 & 95.0 & 93.6 & 94.3\\
\cite{MamanBermano22_UnalignedAMT_ICML} \texttt{SySc} & 86.7 & 79.7 & 82.2 & 82.8 & 91.8 & 86.6 \\

\hline
\end{tabular}
\caption{
Guitar transcription evaluation on the GuitarSet and GAPS datasets. We compare models pre-trained on synthetic data (\texttt{Sy}) and MAESTRO (\texttt{Kg}), trained on GuitarSet (\texttt{Gs}) and GAPS (\texttt{Gp}) using histograms from one-minute windows (\texttt{60s}) and entire tracks (\texttt{F/T}). See text for details.
}
\label{table:guitar_set}
\end{table}
\begin{table*}[t!]
\centering
\setlength{\tabcolsep}{6pt}
\begin{tabular}{lccc|ccc|ccc|ccc}
\toprule
\textbf{Model} & \multicolumn{3}{c|}{\textbf{MAESTRO}} & \multicolumn{3}{c|}{\textbf{GuitarSet}} & \multicolumn{3}{c|}{\textbf{URMP}} & \multicolumn{3}{c}{\textbf{URMP (Histog.)}} \\
\cmidrule(r){2-4} \cmidrule(r){5-7} \cmidrule(r){8-10} \cmidrule(r){11-13}
 & \textbf{P} & \textbf{R} & \textbf{F} & \textbf{P} & \textbf{R} & \textbf{F} & \textbf{P} & \textbf{R} & \textbf{F} & \textbf{P} & \textbf{R} & \textbf{F} \\
\midrule
\multicolumn{13}{c}{Pre-trained Model} \\
\hline
\texttt{Sy}              & 88.3  & 81.6  & 84.6  & 57.9  & 80.7  & 66.2  & 76.2  & 65.4  & 70.1  & 91.8 & 79.8 & 84.9 \\
\hline
\multicolumn{13}{c}{Histogram Supervision MusicNet Piano (ours)} \\
\hline
\texttt{30s}  & 93.0  & 88.2  & 90.4  & 77.8  & 82.5  & 79.4  & 70.1  & 79.6  & 74.5  & 80.1 & 90.8 & 85.0 \\
\texttt{F/T}         & 92.1  & 85.8  & 88.7  & 81.2  & 80.1  & 79.8  & 77.3  & 75.1  & 76.1  & 89.7 & 87.1 & 88.3 \\
\hline
\multicolumn{13}{c}{Histogram Supervision MusicNet Full (ours)}\\
\hline
\texttt{32ms}    & 77.1     & 12.1     & 16.7     & 85.5     & 5.0     & 8.6     & 56.9     & 1.5     & 2.8   & 100.0 & 19.0 & 36.0 \\
\texttt{100ms} & 94.7     & 33.9     & 43.9     & 91.3     & 31.9     & 40.6     & 90.2     & 6.0     & 11.2   & 100.0 & 6.6 & 12.1 \\
\texttt{500ms} & 92.4     & 80.5     & 85.8    & 90.5     & 69.2     & 75.8     & 82.9     & 70.6    & 76.1   & 97.7 & 83.2 & 89.8 \\
\texttt{30s}       & 94.5     & 86.0     & 89.9     & 88.5     & 75.4     & 80.3     & 82.2     & 79.9     & 80.9   & 93.0 & 90.4 & 91.6 \\
\texttt{60s}    & 93.1     & 86.1     & 89.3     & 86.7     & 78.5     & 81.5     & 81.9     & 79.7     & 80.7   & 92.6 & 90.3 & 91.3 \\
\texttt{F/T} & 92.4     & 85.0     & 88.4     & 82.8     & 82.4     & 82.0     & 81.6     & 78.2     & 79.7   & 92.3 & 88.8 & 90.3 \\
\hline
\multicolumn{13}{c}{DTW + Refinement}\\
\hline
\cite{MamanBermano22_UnalignedAMT_ICML} \texttt{AlPl} & 92.6 & 87.2 & 89.7 & 86.6     &  80.4     & 82.9     & 81.7     & 77.6     & 79.6  & 95.6 & 91.0 & 93.2 \\
\cite{MamanBermano22_UnalignedAMT_ICML} \texttt{Al} & 96.4 & 83.4 & 89.2 & 89.0 & 76.9 & 81.5 & 84.0 & 75.2 & 79.3 & 96.6 & 86.8 & 91.3 \\
\bottomrule
\end{tabular}
\caption{
Cross-dataset evaluation.
Training was performed on MusicNet, with evaluation on MAESTRO, GuitarSet, and URMP. For URMP, we also report F-histogram, which does not enforce the 50ms onset threshold.
}
\label{table:musicnet}
\end{table*}
\subsection{Multi-Instrument Transcription}\label{sec:musicnet}
As a final, more challenging, and less controlled experiment, we evaluate the generalizability of the CountEM approach by applying our method to multi-instrument transcription using the MusicNet dataset~\cite{ThickstunHK17_MusicNet_ICLR}, which features recordings of both solo and ensemble performances across various instruments. Unlike the MAESTRO dataset, MusicNet lacks full supervision, as its note labels were derived from aligning audio and MIDI files from different sources, introducing errors, particularly in onset timing~\cite{HawthorneESRSRE18_OnsetsFrames_ISMIR, HawthorneSRSHDE19_MAESTRO_ICLR, MamanBermano22_UnalignedAMT_ICML}. However, a key advantage is that the musical structure was manually verified, ensuring consistency across performances. While fine-grained alignment remains imprecise, note histograms provide a stable and reliable signal, making this dataset well-suited for evaluating our histogram-based supervision approach in real-world, less curated conditions.

Another strength of MusicNet is its diversity in acoustics and instrumentation, making it well-suited for generalization across different musical contexts (zero-shot transcription).

We derive note histograms over entire tracks from unaligned labels. To obtain histograms over shorter chunks, we use loose alignment only to coherently subdivide audio and weakly-aligned labels. Minor errors in onset timing have little impact on histograms computed over 30- or 60-second windows. Future work could explore alternative segmentation techniques for further refinement.

Note that while refined versions for the dataset exist~\cite{MamanBermano22_UnalignedAMT_ICML},
to demonstrate the efficacy of our approach we use the original, weakly-aligned labels.

We also note that we use the MusicNet dataset exclusively for training, as it lacks precise and reliable reference annotations. For evaluation, we again use the MAESTRO and GuitarSet datasets, along with the URMP dataset~\cite{LiLDDS19_MultitrackDataset_TMM}, which consists of string and wind instruments. In URMP the recordings are multi-tracked, where each track is monophonic, making annotations more accurate and reliable. 
While these labels are generally accurate, they are not perfectly precise~\cite{GardnerSMHE22_MultiTaskTranscription_ICLR}. To account for potential timing inaccuracies, we report both the standard 50ms onset F-score, and a high-tolerance metric, referred to as onset \emph{F-histogram}. It is computed similarly to the F-score, but without the 50ms threshold. It compares the sets without considering timing, and serves as an upper bound in cases of annotation errors in onset timing.

We experimented with both pre-trained models appearing in Section~\ref{sec:guitar}, however, the synthetic pre-trained model (\texttt{Sy}) performed better than the piano pre-trained one (\texttt{Kg}). We postulate this is thanks to the diversity in the data used to pre-train \texttt{Sy} (despite being synthetic). Therefore, presented results are from \texttt{Sy}, also used by Maman and Bermano~\cite{MamanBermano22_UnalignedAMT_ICML}, but fine-tuned on MusicNet with our approach. 

As shown in Table \ref{table:musicnet}, our approach improves over the synthetic baseline, even with full-track histograms (\texttt{F/T}), increasing accuracy on MAESTRO from 84.6\% to 88.7\%, and reaching 90.4\% for half-minute segments. It slightly outperforms
a model
from previous work
trained with alignment and pseudo-labels (\cite{MamanBermano22_UnalignedAMT_ICML} \texttt{AlPl}) while relying on a much simpler label estimation method. Notably, our results even with full-track histograms match results using DTW and local-max refinement (\cite{MamanBermano22_UnalignedAMT_ICML} \texttt{Al}), suggesting that DTW may not be essential for this task.

Lastly, we note that when reducing the window size below \texttt{100ms}, accuracy drastically drops, contrary to the MAESTRO dataset where a single frame (corresponding to full supervision) provides best results. This demonstrates that the MusicNet labels contain errors in onset timing, and also shows that our approach can overcome them, as illustrated in Figure~\ref{fig:labeling}.

\section{Conclusion}\label{sec:conclusion}
In this work, we introduced
CountEM,
a novel framework for AMT that leverages histogram-based supervision to eliminate the need for explicit temporal alignment. By replacing traditional alignment strategies with a simple peak-picking mechanism,
CountEM reduces
computational overhead while improving flexibility. Extensive experiments
across piano, guitar, and multi-instrument datasets demonstrated its robustness, achieving performance comparable to or surpassing existing weakly-supervised methods with a significantly simplified label estimation process.

Looking ahead, CountEM's principles could extend to tasks such as instrument recognition, rhythm analysis, and lyrics transcription, particularly in complex polyphonic settings. Further exploration of weakly- and semi-supervised learning strategies could enhance transcription accuracy while minimizing annotation costs. By shifting towards more efficient and scalable supervision mechanisms, CountEM paves the way for data-efficient approaches to music transcription across diverse musical contexts.

\section{Acknowledgements}
This research was supported in part by the Len Blavatnik and the Blavatnik family foundation and ISF grant number 1337/22. This work was funded by the Deutsche Forschungsgemeinschaft (DFG, German Research Foundation) under Grant No. 500643750 (MU 2686/15-1). The International Audio Laboratories Erlangen are a joint institution of the Friedrich-Alexander-Universität Erlangen-Nürnberg (FAU) and Fraunhofer Institute for Integrated Circuits IIS.
\bibliography{referencesMusic, referencesNew}
\end{document}